\documentclass{ws-procs975x65}

\def\prd{Phys.~Rev.~D}        
\def\apjl{ApJ}                
\def\mnras{MNRAS}             
\def\apj{ApJ}                 

\begin{document}

\title{GRAVITATIONAL RADIATION FROM ACCRETING MILLISECOND PULSARS}

\author{MATTHIAS VIGELIUS}
\author{DONALD PAYNE}
\author{ANDREW MELATOS}

\address{School of Physics,\\
University of Melbourne, \\
Parkville, VIC 3010, Australia\\
\email{mvigeliu@physics.unimelb.edu.au}}

\begin{abstract}
It is widely assumed that the observed reduction of the
magnetic field of millisecond pulsars can be connected to the
accretion phase during which the pulsar is spun up by mass accretion
from a companion. A wide variety of reduction mechanisms have been
proposed, including the burial of the field by a magnetic mountain,
formed when the accreted matter is
confined to the poles by the tension of the stellar magnetic field. A
magnetic mountain effectively screens the magnetic
dipole moment. On the other hand, observational data suggests that
accreting neutron stars are sources of gravitational waves, and magnetic
mountains are a natural source of a time-dependent quadrupole moment. We
show that the emission is sufficiently strong to be detectable by
current and next generation long-baseline interferometers. Preliminary
results from fully three-dimensional magnetohydrodynamic (MHD)
simulations are presented. We find that the initial axisymmetric state
relaxes into a nearly axisymmetric configuration via toroidal magnetic
modes. A substantial quadrupole moment is still present in the final
state, which is stable (in ideal MHD) yet highly distorted.
\end{abstract}

\bodymatter

\section{Introduction}
\label{intro}
Despite considerable effort, an unequivocal direct detection
of gravitational waves (GW) is yet to be achieved. The expected wave strain
is several orders of magnitude weaker than the sensitivity
of current interferometric detectors\cite{Lobo03}. One
possibility is to coherently integrate
the signal of a continuous source. In this case, the signal-to-noise
ratio increases with the square root of the observation time\cite{JKSI}.
A variety of physical mechanisms for the generation of continuous
gravitational waves have been proposed\cite{scox1, Owen06}, among them nonaxisymmetric
distortions of the neutron star crust, either due to
temperature variations\cite{Bildsten98, Ushomirsky00} or strong
magnetic fields\cite{Cutler02}, r-mode instabilities \cite{Ster03,
Owen98, Andersson99, Nayyar06}, or free precession \cite{Jones02, vandenbroeck05}.


A promising GW source was recently suggested by
two of us\cite{Melatos05}. Matter accreting onto a neutron star in a
low-mass X-ray binary (LMXB) accumulates at
the magnetic poles until the latitudinal pressure gradient overcomes
the magnetic tension and the plasma spreads equatorwards. The
frozen-in magnetic field is carried along with the spreading matter,
and is therefore compressed, until the magnetic tension is again able
to counterbalance the thermal pressure. This configuration is termed a magnetic
mountain\cite{Payne04}.
During this process, the magnetic dipole moment decreases with
accreted mass, consistent with observational data.\cite{Payne04, Melatos01}

We examine the prospect of detection of GW from magnetic mountains in
section \ref{sec:gr}. In section \ref{sec:st}, we present preliminary
results from fully three-dimensional magneto-hydrodynamic (MHD)
simulations to test the mountains stability.

\section{Gravitational radiation}
\label{sec:gr}
A typical mountain with $M_a \approx 10^{-4} M_\odot$ and
pre-accretion dipolar magnetic field of $B = 10^{12}$ G can provide\cite{Melatos05} a
gravitational ellipticity $\epsilon=|I_1-I_3|/I_1\approx 10^{-5}$,
where $I_1$ and $I_3$ are the principal moments of inerta. $\epsilon$
is considerably higher than the deformation a conventional
neutron star could sustain ($\epsilon \approx 10^{-7}$) via its free elastic response
and is only surpassed by exotic solid strange stars\cite{Owen05}  ($\epsilon \approx 10^{-4}$).

The characteristic GW strain\cite{Brady98} is defined as
$h_c=(128\pi^4/15)^{1/2} G I_{zz} f^2 \epsilon/(dc^4)$, where
$I_{zz}\approx 10^{45}$g cm$^2$ is the principal moment of 
inertia, $f$ the spin frequency, and $d$ the distance to the
object. Fig. \ref{fig} (left) shows $h_c$ as a function of $f$ at a
distance of $d=1$kpc and a mountain
mass of $10^{-8} M_\odot \le M_a \le 10^{-2} M_\odot$ together with
the design sensitivities of LIGO and advanced LIGO for a coherent
integration time\footnote{This is currently too optimistic due to
  computational limitations. The S2 run managed to reduce a five hour stream of
  data\cite{scox1}. Improvements are expected using added computational resources and
  hierarchical search strategies.} of $10^7$ s. A mountain with $M_a
\approx 10^{-6} M_\odot$ should be detectable by LIGO for $f>200$ Hz. This is also
consistent with the observed cutoff at $\sim 700$ Hz in the spin frequency
distribution of LMXBs ---  much slower than the breakup frequency\cite{Bildsten98}.

\def\figsubcap#1{\par\noindent\centering\footnotesize(#1)}
\begin{figure}[b]%
\begin{center}
  \parbox{2.1in}{\epsfig{figure=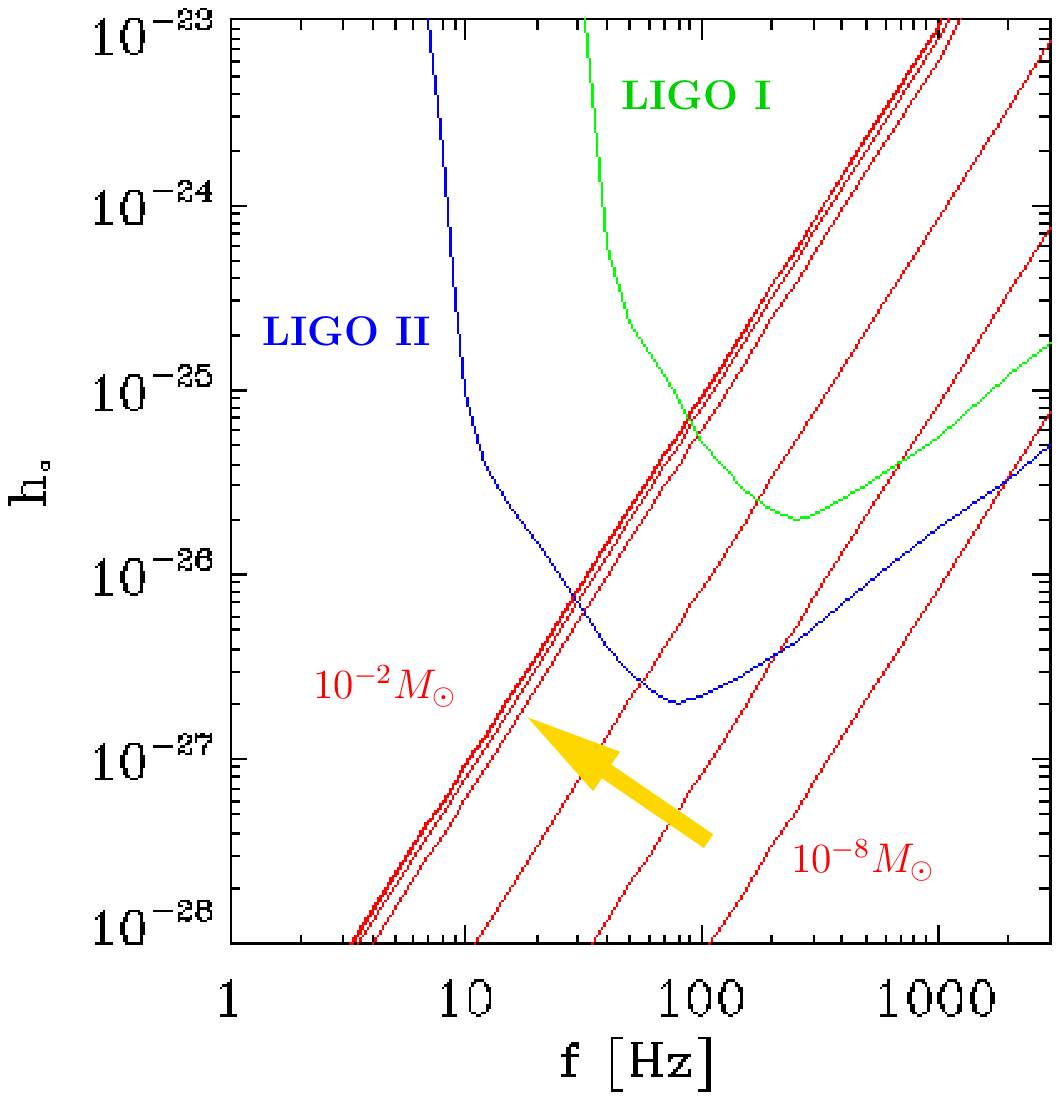,width=2in}\figsubcap{a}}
  \hspace*{4pt}
  \parbox{2.1in}{\epsfig{figure=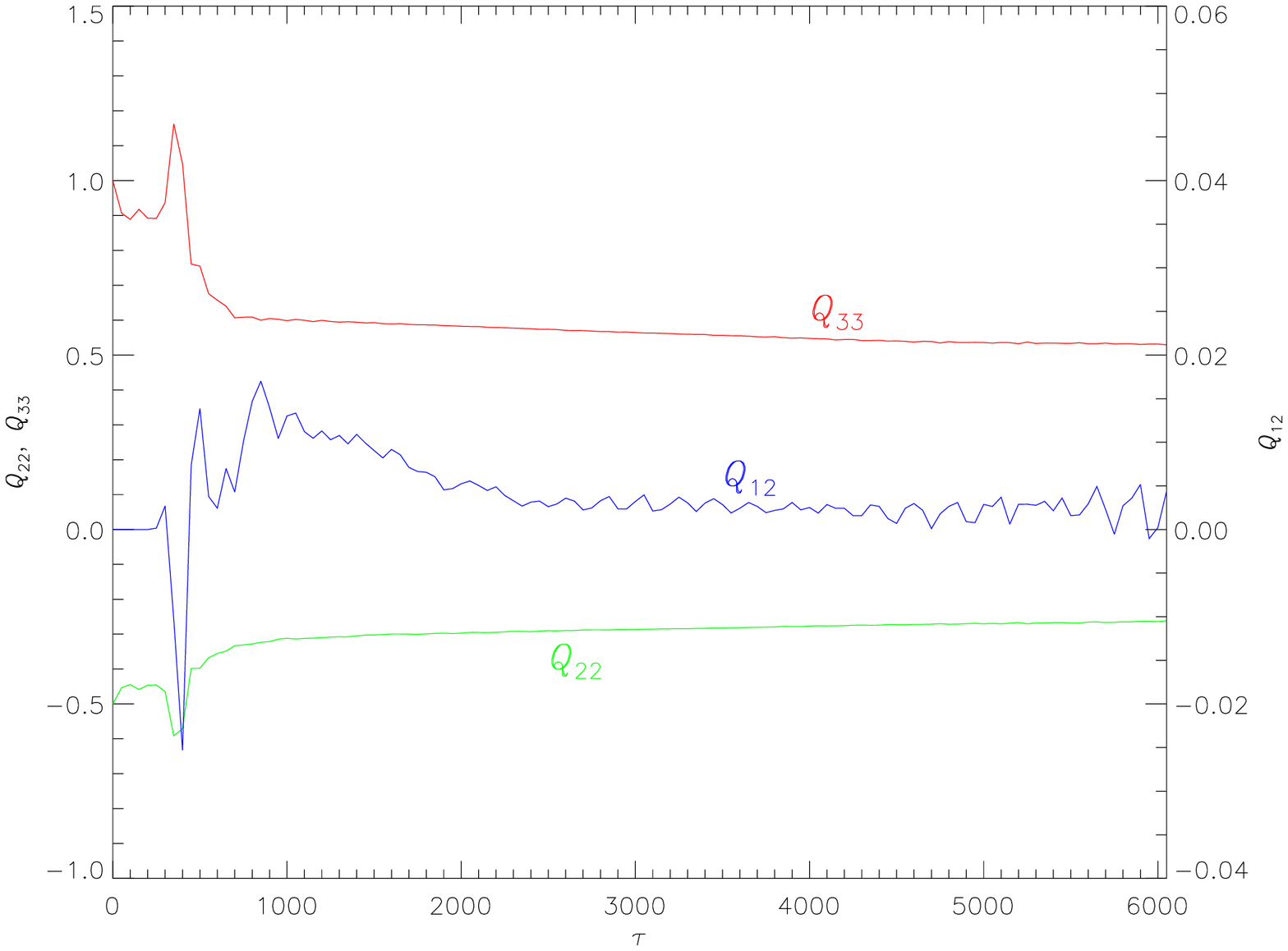,width=2.5in}\figsubcap{b}}
  \caption{(a) Characteristic wave strain $h_c$ for mountain masses
    $10^{-8} M_\odot \le M_a \le 10^{-2} M_\odot$ along with the
    design sensitivity of LIGO and Advanced LIGO for a 99 per cent
    confidence and an integration time of $10^7$ s. (b) Time evolution
    of the quadrupole tensor. The left and righ axes scale the
    diagonal and off-diagonal components, respectively. The time base
    is the radial Alfv\'{e}n crossing time $\tau_\textrm{A}=5.4 \times 10^{-7}$ s.}%
  \label{fig}
\end{center}
\end{figure}

\section{Three-dimensional hydromagnetic stability}
\label{sec:st}
Surprisingly, the distorted magnetic configuration is stable to
axisymmetric modes\cite{Payne06}. However, the full three-dimensional
stability is yet to be examined. We perform
three-dimensional simulations by loading the axisymmetric
configuration into the ideal MHD-code {\sc zeus-mp}. A preliminary
result is displayed in Fig. \ref{fig} (right). Shown is the time
evolution of  the three cartesian quadrupole moments $Q_{22}, Q_{33}$, and
$Q_{12}$, defined as $Q_{ij}=\int d^3 x' \, (3 x_i' x_j'-r'^2
\delta_{ij}) \rho(\mathbf{x'})$, where $\rho$ denotes the
density. After readjusting initially, the system settles down
into a state that still has a sizeable quadrupole
moment. Furthermore, the small magnitude of the off-diagonal element
$Q_{12}$ suggests that the mountain is still nearly (within 
$\approx$1 \%) symmetric about the magnetic axis.

We tentatively interpret these results as a preliminary proof of
three-dimensional stability. However, the
influence of resistivity still needs to be examined.  Resistive
ballooning and resistive Rayleigh-Taylor modes may allow plasma
slippage on a short timescale\cite{Arons}. Non-ideal MHD 
simulations to investigate these effects are currently under way.

\bibliographystyle{ws-procs975x65}

\end{document}